\documentclass[twocolumn, aps, prl, superscriptaddress,floatfix]{revtex4-1}

\usepackage{multirow}
\usepackage{amsmath,amssymb}
\usepackage{graphicx,bm}
\usepackage{slashed}
\usepackage{epstopdf}
\usepackage{ulem} 
\usepackage[usenames]{color}
\usepackage{float}
\usepackage{braket}
\usepackage{appendix}

\renewcommand\sout{\bgroup \color{red} \ULdepth=-.5ex \ULset}

\begin{document}

\title{Algebraic approach to quarkyoniclike configuration and stable diquarks in dense matter  }

\author{Aaron Park}
\email{aaron.park@yonsei.ac.kr}\affiliation{Department of Physics and Institute of Physics and Applied Physics, Yonsei University, Seoul 03722, Korea}
\author{Su Houng~Lee}\email{suhoung@yonsei.ac.kr}\affiliation{Department of Physics and Institute of Physics and Applied Physics, Yonsei University, Seoul 03722, Korea}
\date{\today}
\begin{abstract}
We study the color-spin interaction energy of a quark, a diquark and a baryon with their surrounding baryons and/or quark matter.
This is accomplished by classifying all possible flavor and spin states of the resulting multiquark configuration in both the flavor SU(2) and SU(3) symmetric cases.
We find that while the baryon has the lowest interaction energy when there is only a single surrounding baryon, the quark has the lowest interaction energy when the  surrounding has more than three baryons or becomes a quark gas.  As the short range nucleon-nucleon interactions are dominated by the color-spin interactions, our finding suggests that the baryon modes near other baryons are suppressed due to larger repulsive energy compared to that of a quark and thus provides a quark model  basis for the quarkyoniclike phase in dense matter.  At the same time, when the internal interactions are taken into account, and the matter density is high so that the color-spin interaction becomes the dominant interaction, the diquark becomes the lowest energy configuration and will thus appear in both the dense baryonic and/or quark matter.
\end{abstract}

\maketitle

{\bf Introduction:}
\label{Introduction}
Recent inputs from  multi-messenger astrophysics on neutron star equation of state suggest that there is a sudden increase in pressure as one approaches the core of the neutron star\cite{Kojo:2021wax}. Such behaviour prompted
the appearance of the quarkyonic matter type of equation of state(EOS) for the nuclear matter when the baryon density increases to several times that of the nuclear saturation density\cite{McLerran:2007qj,McLerran:2018hbz, Jeong:2019lhv, Sen:2020peq, Duarte:2020xsp, Duarte:2020kvi, Zhao:2020dvu, Sen:2020qcd, Margueron:2021dtx}.

Recently, using the quark model with color-spin interaction,  we have shown that  the short distance repulsion between the quark and baryon  is smaller than that between two baryons in the lowest energy channel, and that such ingredients naturally lead to quarkyoniclike picture for the EOS when the baryon density reaches four to five times nuclear matter density\cite{Park:2021hqb}.
The relevance of color-spin interaction to describe the dynamics at short distance can be verified in several contexts. In  a previous publication\cite{Park:2019bsz}, we showed that the short distance part of the baryon-baryon interactions for various quantum numbers from the recent lattice calculation\cite{Ishii:2006ec,Inoue:2010hs} can be well reproduced using a constituent quark model with color-spin interaction. The importance of color-spin interaction in nucleon-nucleon repulsion was observed earlier within the quark cluster model\cite{Oka:2000wj}.
The mass splitting between hadrons with different spin orientations, such as the delta and the nucleon  or the psuedoscalar and vector meson, are due to the color-spin interaction\cite{Lee:2007tn}.  Also studies on possible multiquark configurations are  based on the relative strength of the color-spin interaction of the multiquark system compared to that of its lowest hadron threshold\cite{Park:2017mdp,Noh:2021lqs}.

Therefore, when the quark density becomes large so that the baryons start to overlap, the color-spin interaction together with proper application of Pauli principle should be the important dynamics that determines the properties of the  dense matter.   In this work, we will use the quark model with color-spin interaction to study the energy of a quark,  diquarks and a baryon in the dense matter composed of baryons and in the quark matter.
This is accomplished by classifying all possible flavor and spin states of the resulting multiquark configurations in both the flavor SU(2) and SU(3) symmetric cases, and studying the color-spin interactions of these configurations.

We find that while the baryon has the lowest interaction energy when there is only a single surrounding baryon, the quark has the lowest interaction energy when the  surrounding is composed of more than three baryons or becomes a quark gas.
This is an improvement over our precious calculation\cite{Park:2021hqb} as all possible configurations are considered and the surroundings are generalized to study cases with more than one baryons and free quarks.
Our finding implies that the baryon modes near other baryons are suppressed due to larger repulsive energy compared to that of a quark, which is  a property when implemented into phenomenoligical EOS's composed of quarks and baryons leads  to the appearance of  quarkyoniclike phase in dense matter\cite{Jeong:2019lhv,Sen:2020peq,Park:2021hqb}.
 At the same time, when the internal interactions are taken into account, and the matter density is high, similar to the quark density inside a baryon so that the color-spin interaction becomes the dominant interaction, the diquark becomes the lowest energy configuration and will thus appear in both the dense baryonic and/or quark matter.

{\bf Color-spin interaction:}
In the flavor SU(3) symmetric case, the color-spin interaction factor is determined by the following form.
\begin{align}
  H_{CS}&=-\sum_{i<j}^n \lambda^c_i \lambda^c_j  \sigma_i \cdot \sigma_j \label{color-spin0} \\
  &= n(n-10) + \frac{4}{3}S(S+1)+4C_F+2C_C, \label{color-spin}\\
  4C_F &= \frac{4}{3}(p_1^2+p_2^2+3p_1+3p_2+p_1 p_2),
\label{CSF-1}
\end{align}
where $\lambda^c_i$ is the color SU(3) Gell-Mann matrices, $C_F$ is the first kind of the Casimir operator of the flavor SU(3) and $p_i$ is the number of columns containing $i$ boxes in a column in Young diagram. If there is no strange quarks, Eq. (\ref{color-spin}) reduces to the following formula.
\begin{align}
  H_{CS} = \frac{4}{3}n(n-6) + \frac{4}{3}S(S+1)+ 4I(I+1) + 2C_C,
\label{CSF-2}
\end{align}
where $I$ is the total isospin.

In the quark model for a hadron, there will be an extra spatial potential that depends on the distance between two particles multiplying the individual  color-spin factor in Eq~\eqref{color-spin0} so that when calculating the total contribution of the relevant interactions to the hadron mass, the corresponding spatial expectation value has to be taken into account for each pair.  On the other hand, assuming that all the quarks occupy similar spatial dimension and/or the quarks are uniformly distributed, the spatial parts will be universal for all quark pairs.  Taking into account the additional factor that is inversely proportional to the  quark masses of the pair one finds that, to fit the delta-nucleon mass difference, the  overall constant  should be around $C_B/m_u^2 \sim$  18 MeV\cite{Lee:2007tn}, which should be multiplied to the factors in this work to estimates their approximate magnitude when the quark density is similar to that inside a nucleon.   Therefore, to study the relative stability of different configurations, assuming that the density of quarks are uniform in the flavor symmetric limit, one can just compare the color-spin matrix elements given in Eq.~\eqref{color-spin}.

Here, we calculate the color-spin factor experienced by a probe when the surroundings are $n$ baryons, with $n=1,2,3$, or a quark gas, all at a constant baryon density.   The probe will be a quark, a diquark, or a baryon.  As for the diquark, we only consider the most attractive diquark where the  flavor and spin are both antisymmetric.  The
color of the quark and diquarks are in the triplet or antitriplet states, respectively, for which the free energy will be infinity in the confining phase where the surroundings are baryons.  Hence, we also consider the color singlet  three diquark state in flavor SU(3), whose quantum number is the same as the H dibaryon\cite{Jaffe:1976yi}.

{\bf Flavor, color and spin states of a multiquark system:}
Let us consider the multiquark system consisting of $n$ baryons with a probe.
In this work, we assume that the orbital part of the wave function is totally symmetric and the surrounding baryons are in flavor octet states. Since a baryon is a color singlet, the color state of $3n+n_p$ quarks, where $n_p$ is the number of quark in the probe, should be in the color state of the probe. For example, if the probe is a baryon and $n=1$, the flavor-spin coupling state of six quarks should be the conjugate of the color singlet state to satisfy the Pauli exclusion principle. \\

Color : $\begin{tabular}{|c|c|}
  \cline{1-2}
  \quad \quad & \quad \quad   \\
  \cline{1-2}
  \quad \quad & \quad \quad \\
  \cline{1-2}
  \quad \quad & \quad \quad \\
  \cline{1-2}
\end{tabular}$, \quad
Flavor $\otimes$ Spin : $\begin{tabular}{|c|c|c|}
  \cline{1-3}
  \quad \quad & \quad \quad & \quad \quad \\
  \cline{1-3}
  \quad \quad & \quad \quad & \quad \quad \\
  \cline{1-3}
\end{tabular}$\\Then, we can decompose it into the flavor and the spin state as follows.
\begin{align}
  [3,3]_{FS} =& [6]_F \otimes [3,3]_S + [5,1]_F \otimes [4,2]_S + [4,2]_F \otimes [5,1]_S \nonumber \\
  & + [4,2]_F \otimes[3,3]_S + [4,1,1]_F \otimes [4,2]_S \nonumber\\
  & + [3,3]_F \otimes [6]_S + [3,3]_F \otimes [4,2]_S \nonumber\\
  & + [3,2,1]_F \otimes [5,1]_F + [3,2,1]_F \otimes [4,2]_S \nonumber\\
  & + [2,2,2]_F \otimes [3,3]_S. \label{fs-1b1b}
\end{align}
Independently, we can determine the possible flavor states of multiquark system using the outer product.\\

Flavor states of 2-baryon :
\begin{align}
  \mathbf{8}\times \mathbf{8} = \mathbf{1}+\mathbf{8}_{(m=2)}+\mathbf{10}+\overline{\mathbf{10}}+\mathbf{27}.  \label{f-1b1b}
\end{align}
Combining Eq.~(\ref{fs-1b1b}) and Eq.~(\ref{f-1b1b}), we can determine the possible flavor and spin states of six quarks. The allowed flavor and spin states for all cases that we consider are summarized in Table \ref{PFS}.
\begin{table}
  \begin{tabular}{|c|c|}
    \hline
    1b+1b & $\mathbf{1}(S=0), \mathbf{8}(S=1), \mathbf{10}(S=1), \overline{\mathbf{10}}(S=1), \mathbf{27}(S=0)$ \\
    \hline
    \multirow{2}{*}{2b+1b} & $\mathbf{1}(S=\frac{3}{2}), \mathbf{8}(S=\frac{1}{2},\frac{3}{2}), \mathbf{10}(S=\frac{3}{2}), \overline{\mathbf{10}}(S=\frac{3}{2}),$ \\
    & $\mathbf{27}(S=\frac{1}{2},\frac{3}{2}), \mathbf{35}(S=\frac{1}{2}), \overline{\mathbf{35}}(S=\frac{1}{2}), \mathbf{64}(S=\frac{3}{2})$ \\
    \hline
    \multirow{2}{*}{3b+1b} & $\mathbf{1}(S=0), \mathbf{8}(S=1,2), \mathbf{10}(S=1), \overline{\mathbf{10}}(S=1),$ \\
    & $\mathbf{27}(S=0,2), \overline{\mathbf{35}}(S=1), \overline{\mathbf{28}}(S=0)$ \\
    \hline
    1b+1q & $\mathbf{3}(S=0,1), \overline{\mathbf{6}}(S=0,1), \mathbf{15}(S=0,1)$ \\
    \hline
    \multirow{2}{*}{2b+1q} & $\mathbf{3}(S=\frac{1}{2},\frac{3}{2}), \overline{\mathbf{6}}(S=\frac{1}{2},\frac{3}{2}), \mathbf{15}(S=\frac{1}{2},\frac{3}{2}),$ \\
    & $\mathbf{15'}(S=\frac{1}{2},\frac{3}{2}), \mathbf{24}(S=\frac{1}{2},\frac{3}{2}), \mathbf{42}(S=\frac{1}{2},\frac{3}{2})$ \\
    \hline
    \multirow{3}{*}{3b+1q} & $\mathbf{3}(S=0,1,2), \overline{\mathbf{6}}(S=0,1,2), \mathbf{15}(S=0,1,2),$ \\
    & $\mathbf{15'}(S=1), \overline{\mathbf{21}}(S=0), \overline{\mathbf{24}}(S=0,1,2),$ \\
    & $\mathbf{42}(S=0,1,2), \overline{\mathbf{60}}(S=1)$ \\
    \hline
    1b+1d & $\overline{\mathbf{3}}(S=\frac{1}{2}), \mathbf{6}(S=\frac{1}{2}), \overline{\mathbf{15}}(S=\frac{1}{2})$ \\
    \hline
    \multirow{2}{*}{2b+1d} & $\overline{\mathbf{3}}(S=0,1), \mathbf{6}(S=1), \overline{\mathbf{15}}(S=0,1)$ \\
    & $\overline{\mathbf{15'}}(S=1), \overline{\mathbf{24}}(S=0,1), \overline{\mathbf{42}}(S=0,1)$ \\
    \hline
    \multirow{3}{*}{3b+1d} & $\overline{\mathbf{3}}(S=\frac{1}{2},\frac{3}{2}), \mathbf{6}(S=\frac{1}{2},\frac{3}{2}), \overline{\mathbf{15}}(S=\frac{1}{2},\frac{3}{2}),$ \\
    & $\overline{\mathbf{15'}}(S=\frac{1}{2},\frac{3}{2}), \overline{\mathbf{24}}(S=\frac{1}{2},\frac{3}{2}), \overline{\mathbf{42}}(S=0,1),$ \\
    &  $\overline{\mathbf{48}}(S=\frac{1}{2})$ \\
    \hline
    1b+3d & $\mathbf{8}(S=\frac{1}{2})$ \\
    \hline
    2b+3d & $\mathbf{1}(S=0), \mathbf{8}(S=1), \overline{\mathbf{10}}(S=1), \mathbf{10}(S=1), \mathbf{27}(S=0)$ \\
    \hline
    3b+3d & $\mathbf{8}(S=\frac{1}{2}), \overline{\mathbf{10}}(S=\frac{3}{2})$ \\
    \hline
  \end{tabular}
\caption{List of possible flavor and spin states in flavor SU(3) symmetry. b, q and d represent a baryon, a quark and a diquark, respectively. }
\label{PFS}
\end{table}

Now, we investigate the relative magnitude of the interaction which a quark inside the probe sees from the surrounding $n$ baryons using the following formula. A similar formula was used previously by us to compare the  nucleon repulsion with that of the lattice calculation\cite{Park:2019bsz}.

\begin{align}
  \Delta H_{CS}^{n\mathrm{b}+p} &= H_{CS}^{n\mathrm{b}+p} - H_{CS}^{n\mathrm{b}} - H_{CS}^{p},  \label{2b-1} \\
  \Delta H_{CS}^{\mathrm{avg}} &= \frac{1}{n_p n \sum_{C,F,S} d_{CFS}}\sum_{C,F,S}d_{CFS}\Delta H_{CS}^{n\mathrm{b}+p}. \label{2b-2} \\
  d_{CFS} &= d_C d_F d_S m_{FS}.
\end{align}
Here, $n\mathrm{b}$ and $p$ in the superscripts represent $n$ external baryons and the probe, respectively.  The probe will be a baryon, a quark, a diquark or three correlated diquarks.  $n_p$ is the number of quarks in the probe.  We will investigate cases with $n=1,2,3$, and also consider the case where  $n\mathrm{b}$ is replaced by a single quark so as to study the deconfined phase.   $d_C,d_F$ and $d_S$ are the dimensions of the color, flavor and spin states of $3n+n_p$ quarks, respectively, $m_{FS}$ is the multiplicity of the flavor and spin states, and the summation is taken for all possible states.
Here, we divide it by $n_p$ to normalize the result with respect to the single quark case.  We also divide by $n$ to keep the surrounding baryon at constant density for comparison at the same density.

{\bf Free quark gas:}
Finally, we consider the case where the surrounding is a free quark gas. In such a case, we assume that the surrounding free quarks are not correlated with each other, but are correlated with the interacting object to satisfy the Pauli principle.  Therefore, we only need to consider the average value of the color-spin interactions for all possible diquark configurations.
There are four diquark states satisfying the Pauli principle. We represent it for the color SU($N_C$) in the Table \ref{two-quark-interaction}. If we compare it with the results for baryons and a quark, then we should multiply it by 3 to ensure comparison at the same density.

\begin{table}
\begin{center}
\begin{tabular}{c|c|c|c|c}
\hline
\hline
& \multicolumn{4}{c}{$q_i q_j$} \\
\hline
Flavor & $A$ & $S$ & $A$ &$S$ \\
\hline
Color & $A(\overline{3})$ & $A(\overline{3})$  & $S(6)$  & $S(6)$ \\
\hline
Spin & $A(1)$ & $S(3)$ & $S(3)$ & $A(1)$ \\
\hline
$- \lambda_i \lambda_j \sigma_i \cdot \sigma_j$ & $-6-\frac{6}{N_C}$& $2+\frac{2}{N_C}$ &  $-2+\frac{2}{N_C}$ & $6-\frac{6}{N_C}$ \\
\hline
$ \lambda_i \lambda_j $ & $-2-\frac{2}{N_C}$& $-2-\frac{2}{N_C}$ &  $2-\frac{2}{N_C}$ & $2-\frac{2}{N_C}$ \\
\hline
\end{tabular}
\end{center}
\caption{Classification of two quark interaction due to the Pauli exclusion principle. We denote the antisymmetric and symmetric state as $A$ and $S$, respectively. The symbols inside the parenthesis represent the  multiplet state.}
\label{two-quark-interaction}
\end{table}

For a diquark and a free quark, since $\mathbf{3}\times \overline{\mathbf{3}}= \mathbf{1} + \mathbf{8}$, there are two possible color states of three quarks, which will come with the flavor-spin configuration as below.

1. Color : $\begin{tabular}{|c|}
  \hline
  \quad \quad \\
  \hline
  \quad \quad \\
  \hline
  \quad \quad \\
  \hline
\end{tabular}$, \quad
Flavor $\otimes$ Spin : $\begin{tabular}{|c|c|c|}
  \hline
  \quad \quad & \quad \quad & \quad \quad  \\
  \hline
\end{tabular}$\\

2. Color : $\begin{tabular}{|c|c|}
  \hline
  \quad \quad & \quad \quad \\
  \hline
  \quad \quad \\
  \cline{1-1}
\end{tabular}$, \quad
Flavor $\otimes$ Spin : $\begin{tabular}{|c|c|}
  \hline
  \quad \quad & \quad \quad \\
  \hline
  \quad \quad \\
  \cline{1-1}
\end{tabular}$\\
Considering both cases, we can calculate the average value of the color-spin interaction.

For the color-spin interaction between  three correlated diquarks and a free quark, the color and flavor $\otimes$ spin coupling state is as follows.\\
\\
Color : $\begin{tabular}{|c|c|c|}
  \hline
  \quad \quad & \quad \quad & \quad \quad \\
  \hline
  \quad \quad & \quad \quad \\
  \cline{1-2}
  \quad \quad & \quad \quad \\
  \cline{1-2}
\end{tabular}$, \quad
Flavor $\otimes$ Spin : $\begin{tabular}{|c|c|c|}
  \hline
  \quad \quad & \quad \quad & \quad \quad  \\
  \hline
  \quad \quad & \quad \quad & \quad \quad  \\
  \hline
  \quad \quad \\
  \cline{1-1}
\end{tabular}$\\

Flavor states of 7 quarks : $
\mathbf{1}\times \mathbf{3} = \mathbf{3}$ \\

Flavor and spin :
$\begin{tabular}{|c|c|c|}
  \hline
  \quad \quad & \quad \quad & \quad \quad \\
  \hline
  \quad \quad & \quad \quad \\
  \cline{1-2}
  \quad \quad & \quad \quad \\
  \cline{1-2}
  \multicolumn{3}{c}{$\mathbf{3}(S=\frac{1}{2})$}
\end{tabular}$.\\

The $H_{CS}$ factors for this 7 quarks state and the three correlated diquarks are -12 and -24, respectively. In order to compare the result with the interaction factor when one baryon looks at one quark, we need to divide by 2, which corresponds to taking $n_p=6,n=1/3$ in Eq.~(\ref{color-spin}).

For the color-spin interaction between a baryon and a quark, we can use the result in the previous discussions.

{ \bf Color-color interaction:}
The spin independent color-color type of interaction is typically responsible for the confining and Coulomb type of interactions.  While such interactions are important at large separation between color states within a color singlet configuration, the interaction of a colored object with a color singlet configuration is small.  In fact, it is zero if the color singlet configuration has no color polarizations: that is all quarks have the same spatial distribution.   This is so because the color of the immersed object will be the same as the color of the multiquark configuration it makes with the surrounding color singlet baryons.  Consider the following color-color interaction factor composed of $n_p$ quarks from the probe and $3n$ quarks from the $n$ baryons.
\begin{align}
  \sum_{i<j}^{n_p+3n}\lambda_i^c \lambda_j^c & = \frac{1}{2}\bigg[ \big(\sum_i^{n_p+3n} \lambda_i^c \big)^2 - \sum_i^{n_p} (\lambda_i^c)^2- \sum_i^{3n} (\lambda_i^c)^2 \bigg] \nonumber \\
   & = \frac{1}{2}\bigg[ \big(\sum_i^{n_p} \lambda_i^c \big)^2 - \sum_i^{n_p} (\lambda_i^c)^2 \bigg]- \frac{1}{2} \bigg[ \sum_i^{3n} (\lambda_i^c)^2 \bigg] \nonumber\\
   & = \sum_{i<j}^{n_p}\lambda_i^c \lambda_j^c + \sum_{i<j}^{3n}\lambda_i^c \lambda_j^c .
\end{align}
In the second line, the first square bracket is the interaction within the probe while the second that between the color singlet baryons.  That is, the contributions between the quarks in the probe and those in the color singlet configuration cancel out. The result is valid even for color SU($N_C$) with any $N_c$.

{\bf Results:}
\begin{table}
\begin{tabular}{|c|c|c|c|c|}
  \hline
  SU(2)$_F$ &  1b & 2b & 3b & Free quarks \\
  \hline
  quark & 8 & 8.533 & 6.133 & 4.364 \\
  \hline
  diquark & 8 & 8 & 8 & 8 \\
  \hline
  baryon(octet) & 7.111 & 7.111 & 7.111 & 8 \\
  \hline
\end{tabular}
\begin{tabular}{|c|c|c|c|c|}
  \hline
  SU(3)$_F$ &  1b & 2b & 3b & Free quarks \\
  \hline
  quark & 6 & 6.446 & 4.644 & 2.823 \\
  \hline
  diquark & 6 & 6.176 & 5.551 & 6.3 \\
  \hline
  three correlated diquarks & 6 & 6 & 6 & 6 \\
  \hline
  baryon(octet) & 5.714 & 5.78 & 4.944 & 6 \\
  \hline
\end{tabular}\\
\caption{$\Delta H_{CS}^{\mathrm{avg}}$ for different probes (column) in  various surroundings (row) The upper and lower tables are for flavor SU(2) and SU(3), respectively.}
\label{interaction}
\end{table}
As the color-color interactions cancel each other out, to understand the most stable structure at high density where the baryons start to overlap, we consider the color-spin interaction energy of a quark, a diquark or a baryon with the surrounding baryons.
It should be noted that if the immersed object is an isolated colored object, it will require large energy to bring the isolated color charge to the position: this is why the thermal Wilson line is infinity in the confining phase.  Therefore, we also consider the color singlet correlated three diquark state in flavor SU(3) case to search for the minimum energy configuration in the confining phase.  However, when the quark density is large and color confining effects disappear or one is in the quark gas phase, the color of the immersed object can be neglected.

First, we represent $\Delta H_{CS}^{\mathrm{avg}}$ values between the probe and the surrounding for all cases in the Table \ref{interaction}. As can be seen in the table, when the surrounding is a single baryon, the baryon has the smallest interaction energy with the surrounding for both the flavor SU(2) and SU(3) cases. From the values in the Table \ref{interaction}, it can be seen that in the case of flavor SU(3), the magnitude of the color-spin interaction of baryon is smaller than that of others by approximately 5 MeV per single quark if the density is similar to that inside a nucleon so that the previously estimated overall factor is used.  However, when the surrounding becomes  three baryon states or the free quark gas, a quark has the lowest interaction energy. In a series of work\cite{Jeong:2019lhv, Sen:2020peq, Duarte:2020xsp, Duarte:2020kvi}, K.S.J. et al. has shown that a phenomenological model involving quarks and baryons that leads to quarkyonic EOS can be constructed when the short range baryon-baryon repulsion, through the excluded volume effect, are introduced.  More recently, we have shown that in a realistic quark model, the short range quark-baryon repulsion is smaller than the baryon-baryon repulsion in the lowest energy channel, and that such difference leads to quarkyoniclike EOS. Here, we have shown that the quark-baryon repulsion is in general smaller than the baryon-baryon repulsion at high baryon density or free quark gas.  Our results show that the probe that feels the lowest color-spin interaction that dominates the repulsion at short distance is quark rather than baryon, which provides a theoretical justification for quarkyoniclike phase.

Second, to  determine which probe has lower energy in dense matter, one has to consider the internal color-spin factor of the immersed object.   The octet baryon and the diquark both have  color-spin factors  of -8. The additional kinetic energy within the probe typically cancels the color-color interaction strength rendering the color-spin interaction to be the only relevant interaction strength within the probe\cite{Park:2018wjk,Karliner:2017qjm}.
Table \ref{energy} shows the results after considering these internal color-spin factors.
It can be seen that the lowest energy configuration in both the baryon matter and quark matter is the diquark for flavor SU(2) and the color singlet three correlated diquarks for flavor SU(3). Therefore, considering confinement, the color singlet three correlated diquarks is the most stable configuration in dense baryonic matter, when the density becomes similar to the quark density within the baryon so that only the color-spin interaction becomes relevant.  Such configurations will appear in dense baryonic matter. Hence, while H dibaryon do not seem to exist in the vacuum\cite{Park:2016cmg}, similar configurations will appear in dense baryonic matter.  When deconfinement takes place, considering the additional strange quark mass in the color singlet three diquarks, flavor SU(2) diquarks will also appear contributing as important configuration in dense  matter: hence, a diquark matter.
Additionally, if one can show that the interaction between correlated diquarks is relatively more attractive than the corresponding values involving baryons or quarks, then the  so-called diquarkyonic matter consisting of diquarks and baryons or diquark condensation may emerge.

\begin{table}
\begin{tabular}{|c|c|c|c|c|}
  \hline
  SU(2)$_F$ &  1b & 2b & 3b & Free quarks \\
  \hline
  quark & 8 & 8.533 & 6.133 & 4.364 \\
  \hline
  diquark & 4 & 4 & 4 &  4 \\
  \hline
  baryon(octet) & 4.444 & 4.444 & 4.444 & 5.333 \\
  \hline
\end{tabular}\\
\begin{tabular}{|c|c|c|c|c|}
  \hline
  SU(3)$_F$ &  1b & 2b & 3b & Free quarks \\
  \hline
  quark & 6 & 6.446 & 4.644 & 2.823 \\
  \hline
  diquark & 2 & 2.176 & 1.551 & 2.3 \\
  \hline
  three correlated diquarks & 2 & 2 & 2 & 2 \\
  \hline
  baryon(octet) & 3.048 & 3.113 &   2.277 & 3.333 \\
  \hline
\end{tabular}
\caption{Same as Table III after internal interactions within the probes are considered.}
\label{energy}
\end{table}

\section*{Acknowledgments}
This work was supported by Samsung Science and Technology Foundation under Project Number SSTF-BA1901-04. The work of A.P. was supported by the Korea National Research Foundation under the grant number 2021R1I1A1A01043019. The authors would like to thank Kiesang Jeong and Hyungjoo Kim for useful discussions.

\end{document}